\documentclass{iau} 
\usepackage{graphicx}
\usepackage{amsmath}
\usepackage[square,sort]{natbib}

\newcommand{\apj}{ApJ}
\newcommand{\mnras}{MNRAS}
\newcommand{\apjl}{ApJ}
\newcommand{\araa}{ARA\&A}
\newcommand{\aap}{A\&A}

\title[The AGN Jet Model of the Fermi Bubbles] 
{The AGN Jet Model of the Fermi Bubbles}

\author[Fulai Guo]   
{Fulai Guo$^{1}$
}

\affiliation{$^1$Key Laboratory for Research in Galaxies and Cosmology,  \\ Shanghai Astronomical Observatory, Chinese Academy of Sciences, \\ 80 Nandan Road, Shanghai 200030, China \\ email: {\tt fulai@shao.ac.cn} \\[\affilskip]
 }

\pubyear{2016}
\volume{322}  
\jname{The Multi-Messenger Astrophysics of the Galactic Centre}
\editors{S. Longmore, G. Bicknell \& R. Crocker, eds.}
\begin{document}

\maketitle

\begin{abstract}
The nature and origin of the {\it Fermi} bubbles detected in the inner Galaxy remain elusive. In this paper, we briefly discuss some recent theoretical and observational developments, with a focus on the AGN jet model. Analogous to radio lobes observed in massive galaxies, the {\it Fermi} bubbles could be naturally produced by a pair of opposing jets emanating nearly along the Galaxy's rotation axis from the Galactic center. Our two-fluid hydrodynamic simulations reproduce quite well the bubble location and shape, and interface instabilities at the bubble surface could be effectively suppressed by shear viscosity. We briefly comment on some potential issues related to our model, which may lead to future progress.

\end{abstract}

\firstsection 
\section{Introduction}

The  {\it Fermi} bubbles are a large structure recently discovered in the inner Galaxy by the {\it Fermi Gamma-ray Space Telescope} (\citealt{su10}, \citealt{dobler10}, \citealt{ackermann14}). The two gamma-ray bubbles have a bilobular shape, extending to $\sim 50^{\circ}$ above and below the Galactic center (GC). Assuming the distance of the GC to be $d\sim 8.5$ kpc, the major axis of each bubble roughly aligns with the Galaxy's rotation axis, having a length of $\sim d {\rm tan}50^{\circ}\sim10$ kpc. The {\it Fermi} bubbles have a hard $\sim E^{-2}$ spectrum between $1$ GeV and $100$ GeV, and sharp edges. They have counterparts in microwave, previously observed by the {\it Wilkinson Microwave Anisotropy Probe} (WMAP) and referred as the WMAP haze \citep{finkbeiner04a}.

The microwave emission from the {\it Fermi} bubbles is usually considered to arise from synchrotron emission of a hard population of cosmic ray (CR) electrons (\citealt{dobler08}, \citealt{dobler12}), while the origin of their gamma-ray emission remains debated. The leptonic scenario assumes that the gamma ray emission comes from inverse Compton scattering (ICS) of the ambient interstellar radiation field (ISRF) by the same population of hard electrons (\citealt{dobler10}, \citealt{su10},\citealt{guo12a}). Alternatively, in the hadronic scenario, the gamma ray emission results from CR protons, which collide inelastically with the ambient gas and produce neutral pions, which decay into gamma rays (\citealt{crocker11}, \citealt{zubovas11}, \citealt{mou15}). With properly chosen CR spectrum, both the leptonic and hadronic scenarios fit gamma ray data quite well, while the latter seems to require an extra population of primary electrons to explain the observed microwave emission from the {\it Fermi} bubbles \citep{ackermann14}. 

The elusive origin of the {\it Fermi} bubbles remains a topic of active research. \citet{crocker11} suggested that the bubbles are powered by CR protons continuously injected by supernova explosions in the GC over the last few Gyrs. \citet{crocker14} and \citet{crocker15} further argue that the bubbles are inflated by a nuclear outflow driven by GC star formation over the last few 100 million years. 
On the other hand, Sgr A$^{*}$ may be a natural energy source for the {\it Fermi} bubbles. Black hole accretion events often release winds and jets, as observed in quasars and radio galaxies \citep{fabian12}. Observational evidence for recent AGN activity at the GC includes (1) there appear to be two young stellar disks within $0.5$ pc of Sgr A$^{*}$ with typical stellar ages of $6\pm 2$ Myr (\citealt{genzel03}, \citealt{paumard06}), which may be remnants of recent accretion flows onto Sgr A$^{*}$; (2) the H$\alpha$ emission of the Magellanic Stream peaks toward the south Galactic pole, which may be energized by ionizing photons from Sgr A$^{*}$ about $1 - 3$ Myrs ago \citep{bland13}.

The GC AGN activity could potentially inflate bubbles through jets or winds. The AGN jet model for the {\it Fermi} bubbles was first studied by \citet{guo12a}, motivated by the morphological similarity between the {\it Fermi} bubbles and extragalactic radio lobes. The model is further studied by \citet{guo12b}, \citet{yang12}, and \citet{yang13}, arguing that the bubbles were produced by a recent AGN jet activity about few Myrs ago, and the origin of the gamma ray emission is leptonic. In contrast, the AGN wind model typically assumes that the gamma ray emission of the {\it Fermi} bubbles results from CR protons. \citet{zubovas11} proposed that the {\it Fermi} bubbles were inflated by wide-angle, ultrafast outflows from Sgr A$^{*}$ in its quasar phase around 6 Myr ago (also see \citealt{zubovas12}). Alternatively, motivated by recent numerical studies of hot accretion flows onto SMBHs \citep{yuan14}, \citet{mou14} and \citet{mou15} argue that the {\it Fermi} bubbles were continuously inflated by winds from hot accretion flows around Sgr A$^{*}$ during the past $\sim 10$ Myrs.

\section{Overview}
\begin{figure}[htp!]
\begin{center}
 \includegraphics[trim=0 0 0 0.5, clip,scale=0.6]{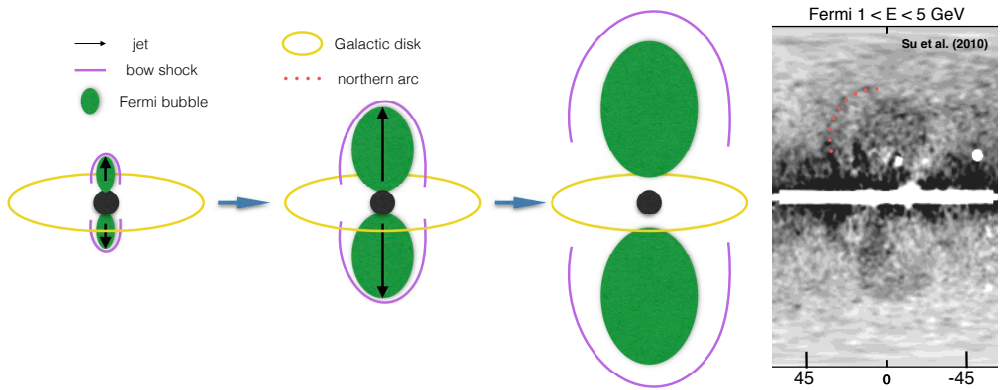} 
 \caption{The AGN jet model of the {\it Fermi} bubbles. The left three sketches describe the formation of the {\it Fermi} bubbles by opposing jets from the Galactic center. The AGN jet activity inflated bilobular bubbles, and produced bow shocks embracing the bubbles, which resemble the northern arc observed by {\it Fermi Gamma-ray Space Telescope}. The rightmost panel, adapted from \citet{su10}, shows the residual {\it Fermi} $1 - 5$ GeV map after subtracting dust and disk templates. The dotted line in the right panel denotes the northern arc described in \citet{su10}.}
   \label{fig1}
\end{center}
\end{figure}

In this section, we discuss the AGN jet model of the {\it Fermi} bubbles developed in \citet{guo12a} and \citet{guo12b}. Our model is motivated by similar opposing jet events observed in massive galaxies \citep{mcnamara12}. AGN jets have been proposed to explain extended extragalactic radio lobes, whose radio emissions are clearly synchrotron emission of CR electrons in magnetic fields \citep{longair73,scheuer74,blandford74}. Observations of radio jets and lobes indicate that AGN jets accelerate and carry CR electrons, producing CR-filled kpc-scale bubbles. CR electrons in AGN bubbles are expect to produce gamma ray emissions through ICS of cosmic microwave background (CMB) photons. However, due to its limited sensitivity and resolution, gamma-ray observation can not easily detect AGN bubbles. The {\it Fermi} bubbles in our Galaxy are probably one of those rare nearby AGN bubbles that can be detected in gamma rays, and similar gamma-ray AGN bubbles may also be present in Centaurus A \citep{sun16}, the Circinus galaxy \citep{hayashida13}, and potentially Andromeda \citep{pshirkov16}.

The process of producing kpc-scale bubbles by AGN jets has been extensively studied by numerical simulations (e.g., \citealt{norman82}). The connection between jet properties and the shapes of resulting bubbles has been investigated numerically by \citet{guo15} and \citet{guo16}. Taking CRs as a second fluid, \citet{guo11} performed numerical simulations of AGN jets composed of both thermal gas and cosmic rays in galaxy clusters. Adopting similar methodology as in \citet{guo11}, \citet{guo12a} performed the first numerical simulation of the AGN jet model for the {\it Fermi} bubbles, showing that a pair of opposing jets originated from Sgr A$^{*}$ could reproduce the {\it Fermi} bubbles quite well with the observed location,size, and shape.

In our model, the jet evolution in the Galactic halo can be described in the left three sketches in Figure 1. The AGN jets were released nearly along the Galaxy's rotation axis about $1 - 3$ Myr ago, and were active for a period of about $0.1 - 0.5$ Myr. Each jet interacts with the halo gas at its head, where the overpressured jet material flows laterally away from the jet head, and subsequently flows backward down to lower latitudes. The jet backflow constitutes the observed sharp edges of the {\it Fermi} bubbles, across which cosmic ray diffusion must be significantly suppressed, possibly due to parallel magnetic field lines draped by the vertically rising bubbles. As implied by the observed smooth edges of the bubbles, interface instabilities (e.g., Kelvin-Helmholtz and Rayleigh-Taylor instabilities) at the bubble surface should be suppressed, potentially by shear viscosity as demonstrated in \citet{guo12b}. 

As shown in Figure 1, the AGN jet event produced a bow shock surrounding the {\it Fermi} bubbles, which is a generic prediction in probably all energy injection models in the GC. Adapted from \citet{su10}, the rightmost panel of Figure 1 shows a part of the residual {\it Fermi} $1 - 5$ GeV map after subtracting dust and disk templates. The dotted line indicates the location of the northern gamma-ray arc described in \citet{su10}, which appears to be associated with the North Polar Spur (NPS) observed in radio and soft X-rays. In our simulations, the location of the bow shock surrounding the current {\it Fermi} bubbles is very close to the northern arc, suggesting that the northern arc (i.e., the NPS) is the bow shock induced by the {\it Fermi} bubble event. Further observational investigations are surely needed to confirm or rule out this scenario for the NPS.

\section{Discussion}

While the AGN jet model for the {\it Fermi} bubbles is promising, there are still some challenging issues specifically related to its current version. The total power of two jets in our fiducial run in \citet{guo12a} is $\sim 0.3$ the Sgr A$^{*}$ Eddington luminosity, which might correspond to an accretion rate too high for hot accretion flows typically hosting AGN jets. However, the jet power could be much less if the halo gas density is lower \citep{guo12b}, and the jet power of hot accretion flows could exceed the accretion rate of rest mass energy \citep{sadowski13}.

The current age of the bubbles in our model is about $1 - 3$ Myr, which is constrained by the cooling time of CR electrons. However, several pieces of evidence seem to suggest that the GC AGN activity occurred a slightly longer time ago. The ages of young stars in the GC stellar disks are about $6\pm 2$ Myr (\citealt{paumard06}). Ultraviolet absorption-line spectra suggests the existence for a biconical outflow from the GC probably driven by the {\it Fermi} bubble event over the past $2.5 - 4$ Myr \citep{fox15}. Observations of soft X-ray emission lines toward the bubble regions also suggest an expansion age of about 4 Myr \citep{miller16}. If the real bubble age is around 4 Myr, to keep their hard spectra CR electrons in the {\it Fermi} bubbles should be continuously reaccelerated by the second order Fermi acceleration \citep{mertsch11} or other mechanisms, unless the gamma ray emission from the bubbles is dominated by CR protons.

\acknowledgments

FG gratefully acknowledges support from Chinese Academy of Sciences through the Hundred Talents Program and National Natural Science Foundation of China (Grant No. 11643001 and 11633006).


\end{document}